\documentstyle[12pt,aasms]{article}
\begin{document}

\title{DYNAMICAL HISTORY OF Ly $\alpha$ CLOUDS}

\author{R. Srianand\\
        Inter-University Centre for Astronomy and Astrophysics \\
        Post Bag 4, Ganeshkhind, Pune 411 007, India,\\
        email anand@iucaa.ernet.in}
\centerline{( To appear in Ap. J)}
\begin{abstract}

The clustering properties of Ly ${\alpha}$ lines are analysed using an
intermediate resolution ($\sim 1\AA$) spectra of 67 QSOs compiled from
the literature. The pair velocity correlation function indicates a weak
excess in the velocity intervals up to ${\rm \Delta v\;=\;600\;
km\;s^{-1}}$. The z-integrated probability distribution of interline
spacings also confirms the excess in the low velocity intervals.  The
dependence of pair velocity correlation on redshift and equivalent
width are investigated. The cross-correlation properties of Ly $\alpha$
clouds with metal line redshifts are analysed. We do not find any
tendency of Ly $\alpha$ lines to cluster around metal line systems.
Depedence of the cross-correlation on redshift and equivalent widths of
Ly $\alpha$ clouds are investigated.  Various implications of the
results are discussed.

\vskip 0.4in

\noindent{{\bf\it Subject headings:} QSO: Ly $\alpha$ absorption
lines- clustering}

\end{abstract}

\section{ INTRODUCTION}

Ly $\alpha$ absorption lines seen in the spectra of QSOs, being the
most abundant objects at higher redshifts, can provide us valuable
information regarding the intergalactic medium at the earlier epochs.
Though the properties of these lines have been studied for the past two
decades, we are still not in a position to understand the origin of
these lines.  Since these lines do not show appreciable amount of
metallicity and clustering they are considered to be primordial
intergalactic material. Sargent et al.(1980) proposed the pressure
confined model for these lines. Though this model explains the observed
properties of Ly $\alpha$ lines the origin of the clouds is not well
defined. The dynamical evolution of Ly $\alpha$ lines in this model
will be mainly due to environmental effects.

Rees (1986), Ikeuchi (1986) and  Bond, Szalay, and Silk (1988),
proposed the gravitationally induced formation scenarios for the origin
of Ly $\alpha$ clouds. Ostriker {\&} Ikeuchi (1983) and Chernomordik
{\&} Ozernoy (1983) proposed shocked shell models in which the clouds
originate in the fragmenting shells, which is a natural extension of
the theory of how some stars form in the interstellar medium. Last two
models  have physical origin for the formation of the Ly $\alpha$ lines
and can predict their spatial distribution and its evolution with
redshift.  Studying the  correlation properties and their evolution can
allow one to understand the dynamical history of the Ly $\alpha$ clouds
and discriminate between various models.

Webb(1987) showed, in the case of Ly ${\alpha}$ absorption lines
obtained using high resolution spectroscopy, that there is an excess in
the pair velocity correlation for scales ${\rm \sim 300\;km\;s^{-1}}$.
Ostriker, Bajtlik and Duncan (1988; OBD here after) showed that the
line interval distribution function is the better tool for studying the
clustering properties than the pair velocity correlation and showed a
significant excess in the lower velocity scales in their low resolution
sample. Recently we (Srianand {\&} Khare 1994) also confirmed the
excess in these velocity scales using high resolution observations of 8
QSOs.  However the distribution of lines, both in the case of low as
well as high resolution samples, seems to be uniform beyond 300 km
s$^{-1}$.

Tytler (1987) based on the single powerlaw neutral hydrogen column
density distribution, seen over six orders of magnitude in column
density, proposed the same origin for metal line systems as well as Ly
$\alpha$ clouds. There are indications, from the HST observations,
that  at least few {\%} of Ly $\alpha$ clouds at very low z may be
associated with galaxies (Lanzetta et al. 1994).  Deep imaging studies
at low redshifts indicate the association between metal line systems
and luminous galaxies (Bergeron (1988); Steidel (1993)). Thus studying
the cross-correlation between the metal line systems and Ly $\alpha$
lines and its evolution with redshift can provide us some useful
information. Barcons and Webb (1990) studied the cross-correlation
properties and did not find any significant excess on any velocity
scale.

In this paper the clustering properties of the Ly $\alpha$ clouds among
themselves and with metal line systems are reanalysed in detail with an
extended intermediate resolution sample compiled from the literature.
The details of the data used  in the analysis are given in section 2.
The results of the auto-correlation analysis and the dependence of
auto-correlation on equivalent width and redshift are presented in
section 3. The results of cross-correlation analysis between Ly
$\alpha$ lines and metal line systems and the dependence of
cross-correlation on  equivalent width and redshift are presented in
section 4. Discussion and results are presented in section 5 and 6
respectively.

\section{ DATA SAMPLE}

Most of the earlier studies of correlation properties of Ly $\alpha$
clouds were performed with small samples obtained with varied
resolution, S/N and line selection techniques. In order to get any
significant result one requires a data sample compiled with large
number of spectra obtained with similar resolution and S/N. Such a
sample was compiled recently by us from the literature for analysing
the properties of Ly $\alpha$ clouds in the vicinity of QSOs (Srianand
and Khare 1995).  Our data sample is obtained, by considering the QSO
spectra observed with spectral resolution resolution $\le \;100\;{\rm
km\;s^{-1}}$, from the literature. In total there are 67 QSO spectra
covering a redshift range between 1.7 and 4.0.  The necessary details
of the data with references are given in Table 1.

${\rm z_{max}}$ is the smaller of observed maximum and emission
redshift of the QSOs and ${\rm z_{min}}$ is the larger of the observed
minimum and the redshift  corresponding to the Ly $\beta$ emission. E
is the minimum detectable equivalent width of a line of 5$\sigma$ level
in each spectra. In most of the cases these values are taken as given
by the authors of the parent references. Whenever these values are not
given E is taken to be five times the largest observable error in the
equivalent width of any line in the relevant redshift range.  ${\rm
z_{abs}}$ is the redshift of known metal line system in the observable
range of Ly $\alpha$ absorption.  In order to avoid any ambiguity
introduced by the different selection criteria used by various authors
the metal line systems are graded as A, B and C,  which are also
provided in the table 1. The criteria used are discussed in section 4.  The
gaps in the spectra, the metal lines and the Ly ${\alpha}$ due to metal
line systems are considered as the excluded regions for the Ly $\alpha$
observations. Since the presence of metal line systems can cause a
spurious signal in the correlation function even very doubtful systems
are excluded from the Ly $\alpha$ sample.  Very strong Ly $\alpha$
lines, ${\rm W > 5\AA}$, are also considered to be excluded regions as
they may well belong to the damped Lyman alpha systems.

\section{ PAIR VELOCITY CORRELATION FUNCTION}

The pair velocity correlation between Ly $\alpha$ clouds along the line
of sight has been one of the tools commonly used to study the
clustering properties of Ly $\alpha$ clouds (Sargent et al. 1980). It
is customary to define the pair velocity correlation at any velocity
interval v, ${\rm \xi (v)}$ the excess probability over the expectation
based on randomly distributed absorbers, as

\begin{equation}
 {\rm \xi (v) = {n_{obs}(v)\over n_{exp}(v)} -1}.
\end{equation}
where ${\rm n_{obs}(v)}$ is the number of observed pairs with velocity
`v' and ${\rm n_{exp}(v)}$ is the expected number of pairs obtained
assuming the cloud distribution to be random. ${\rm n_{exp}}$ is
calculated taking into account the redshift evolution of number density
of Ly $\alpha$ lines, obtained from the sample used here(Srianand {\&}
Khare 1995).  Sargent et al. (1980) used a ramp-shaped function
to account for the limited redshift coverage of each spectra.
Instead of taking any correction function, for each Ly $\alpha$ line
the expected number of lines in various velocity bins are calculated
taking into account the observable range in the particular spectrum.
The errors in the correlation functions are calculated using the
analytic relations given by Mo, Jing and Borner(1992). These relations
provide a realistic estimation of errors for $\xi$, in the case of weak
clustering (i.e. $\xi\;<1$). The observed number of pairs in each
velocity bin with expected number and their $2\sigma$ errors are
plotted in figure 1. (for equivalent width cutoff ${\rm
W_{min}\;=\;0.3\AA}$).  Only lines which are 8 Mpc away from the QSOs
are considered in order to avoid the complications due to proximity
effect of QSOs.

The excess in the low velocity bins is evident from the figure 1. For
the relative velocities between 200 and 400 km s$^{-1}$ the observed
value of ${\xi}$ is ${0.316\pm 0.104}$, a clear 3$\sigma$ excess. The
excess seems to extend up to 600 km s$^{-1}$ with ${\xi =
0.23\pm0.08}$. Note that the deficit seen in the first bin may be an
artifact of the blending and actual clustering may be much stronger
than the one seen here. Also if the Ly $\alpha$ clouds have small
peculiar velocities on top of Hubble flow then the real spatial
correlation may be much stronger than the one implied by the velocity
correlation. In order to confirm this excess the distribution of
interline spacing described by OBD is analysed.

\subsection{ Distribution of interline spacing}

Considering the z-integrated probability distribution of size intervals
between adjacent absorption lines, OBD showed a weak excess in the scales
200-600 km s$^{-1}$. If the line distribution is locally Poisson, then the
expected distribution of line intervals is

\begin{equation}
 {\rm P(x) = \exp^{-x}} .
\end{equation}

Where ${\rm x={ \Delta z/ \overline{\Delta z}}}$, is the line
interval scaled to the local mean. One can approximate in any given
redshift range,

\begin{equation}
 {\rm{1\over \overline{\Delta z}} = N_o(1+z)^{\gamma}}
\end{equation}
where ${\rm N_o\;and\;\gamma}$ are constants obtained from fitting the
observations.  Following OBD, the effect of blending is modeled by
assuming the number of low $\Delta {\rm z}$ intervals is reduced by a
factor,

\begin{equation}
{\rm 1-exp\bigg[-\bigg({\Delta z\over \delta_b}\bigg)\bigg]}
\end{equation}
relative to Poisson, where Gaussian blending scale ${\rm \delta_b}$ is

\begin{equation}
{\delta_b = \zeta {W_{min}(z)\over \lambda_\alpha}(1+z)}
\end{equation}

The redshift density of line intervals and z-integrated probability
distribution for line interval sizes are derived, using the equations
given by Babul(1991), taking into account the excluded regions in the
spectra. Both the distributions are fitted simultaneously to get the
best fit values of $ \zeta $, $\gamma$ and ${\rm N_o}$. The values
obtained for ${\rm w_{min} = 0.3\AA}$ are 1.0, 2.3 and 3.80
respectively. The line interval distribution and the expected
distribution for the best fitted truncated Poisson model are given in
fig 2. Kolmogorov-Smirnoff test shows, see
fig(3), the probability that the maximum deviation between the observed
and the predicted cumulative distributions to occur by chance is
0.0001, and the maximum deviation occurs at the interval separation
0.23. Thus the distribution of interline spacing also confirms the
excess correlation in the low velocity intervals.

\subsection{ Dependence of clustering on rest equivalent width }

 At least 32${\%}$ of the very low redshift Ly $\alpha$ lines seem to be
associated with luminous galaxies (Lanzetta; 1994). It may be possible
that among the Ly $\alpha$ clouds a fraction may be associated with
galaxy kind of objects or their progenitors (say metal line systems at
higher redshifts), and rest of them are due to intergalactic clouds. In
such a scenario one would expect to see strong lines to cluster more
than weak lines as the lines associated with galaxies are expected to
be stronger.

Models involving biasing for the formation of structure in the universe
predict that the objects formed at the sights associated with the
stronger potential wells should be more strongly clustered. If Ly
$\alpha$ clouds are in some way associated with primordial density
fluctuations, and the rest equivalent width is related to cloud mass,
then we might expect the strong absorption lines to show clustering which
is stronger than their weak counterparts.

The pair velocity correlation analysis is performed for different
values of equivalent width cutoff.  The calculated values of $\xi$ for
various low velocity bins are given in Table 2. A glance at the table
reveals that in all cases there is more than 2$\sigma$ excess in the
velocity interval 200-400 km s$^{-1}$ . Also there seems to be a
moderate increase in clustering amplitude with equivalent width
cutoff.  Note that Crotts(1989) also found that the stronger lines tend
to cluster more readily compared to weak lines. In the case of metal
lines also Steidel and Sargent (1992) showed strong Mg II lines
clustered more readily than the weaker ones. Recently Cristiani et al
(1995) showed the correlation function to depend on column density for
the lines in the spectra of QSO 0055-269.

However the increase in $\xi$ is marginal and within $2\sigma$ errors.
Thus our results do not show any statistically significant increase in
$\xi$ with equivalent width cutoff. And even if the increase is real
it is not to the extent that is expected if most of the high equivalent
width Ly $\alpha$ lines are similar to the nonLLS C IV systems. Thus
it seems strong Ly $\alpha$ lines at earlier epochs are not associated
with the regions similar to the present day galaxies.

\centerline{\bf 3.3 Redshift dependence of pair velocity correlation}

Structures formed due to gravitational instabilities and explosions are
predicted to show spatial clustering. However the two scenarios predict
different evolutionary pattern for clustering with time. The sample
used here is divided into two subsamples at z =2.9 (mean redshift of
the sample) and analysed  to see any possible change in the pair velocity
correlation function.  The expected and observed number of pairs for
various velocity bins together with 2$\sigma$ errors are shown in
figure 4a and figure 4b for low and high z respectively (with ${\rm
W_{min}\; = \;0.3\AA}$). While the correlation excess for low z
subsample extend up to 800 km s$^{-1}$ with ${\xi =0.263\pm 0.119}$ the
high z subsample show weaker correlation strength in the same velocity
interval ($\xi = 0.155\pm0.106$). The increase is moderate and is with
in 1$\sigma$ errors in $\xi$. The velocity width of 800 km s$^{-1}$ at
z=2.4 ( average z of low z subsample), for ${\rm q _o}$ = 0.5 and ${\rm
H_o}$ = 100 km s$^{-1}$, corresponds to a distance scale of 0.5 Mpc.
For z = 3.4 the same distance corresponds to a velocity width of 1600
km s$^{-1}$.

The results for high equivalent width lines (with ${\rm W_{min}\; =
\;0.6\AA}$) are shown in figures 4c and 4d. A glance at the figures
reveals a moderate evolution in the correlation function. While
$\xi\;=\; {\rm 0.700 \pm 0.342}$ for velocity interval 200-800 km s
$^{-1}$ for z $<$ 2.9 the higher redshift samples show $\xi\;=\;{\rm
0.080\pm 0.200}$. Also in the lower redshift subsample the excess seems
to extend up to 2500 km s$^{-1}$. This velocity scale corresponds to a
distance of $\sim$ 4.0 Mpc (for ${\rm q_o}$ = 0.5 and ${\rm H_o}$ = 100
km s$^{-1}$ Mpc$^{-1}$), which is roughly the clustering scale of the
present day luminous galaxies.  It seems for the high equivalent width
lines either the spatial correlation strength has increased with time
and/or the systemic velocity of the clouds increased with time.
However, one can see (in figure 4.) a large positive and negative
deviations which are more than 2 $\sigma$ errors in the higher velocity
intervals. Since the number of QSO spectra used in the sub-samples are
small these deviations may very well be noise.  This prevent us from
making any firm conclusions about the clustering scales.  In any case
one would expect to find more correlation power in the pair velocity
correlation for the local Ly $\alpha$ clouds (Ly $\alpha$ lines
observed with HST).

Bahcall et al. (1993) showed, based on HST observations, that there is
no excess in the calculated two point correlation function for
absorption lines in 13 QSO spectra. They noted that with the available
data, the observed Ly $\alpha$ correlation function is not inconsistent
with the hypothesis that Ly $\alpha$ lines are correlated as strongly
as are galaxies.  In order to test this hypothesis at higher
significance level one may require data set that is 3 to 4 times bigger
than that is available in the literature.

A strong prediction of any model of structure formation based on
gravitational instability is that the correlation function should
increase substantially with time. Such a picture also predicts that the
systemic velocities associated with gravitationally bound systems would
also increase with time as they reach virial equilibrium. However the
explotionary theories of structure formation (Vishniac, Ostriker and
Bertschinger 1985; Weinberg,Ostriker and Dekel 1989) suggest either
unchanging or weakening of pair velocity correlation with time. If the
increase in $\xi$ shown by the high equivalent width lines are real, and
confirmed with higher significance it will favour gravitationally
induced structure formation models over the explosion models. Also
difference in clustering strength of metal lines and Ly $\alpha$ lines
requires some sort of biasing in the  usual CDM models (Salmon and
Hogan, 1986).

\section { Ly $\alpha$ - METAL LINE SYSTEM CROSS CORRELATION}

Morris et al (1993) showed, based on very low z observations, an excess
in cross-correlation between Ly ${\alpha}$ and the luminous galaxies,
though not as strong as the Galaxy-Galaxy autocorrelation, in the lower
velocity scales. Studying the cross correlation function at different
epochs will enable us to understand the origin and evolution of Ly
$\alpha$ clouds. There are evidences for most of the heavy element
absorption systems seen in the spectra of QSOs to be  associated with
luminous galaxies (Bergeron, 1988; Steidel, 1993).  Strong two point
velocity correlation also confirms the metal line systems to be the
possible sights of present day luminous galaxies. Barcons and Webb
(1990) studied the cross-correlation between heavy element systems and
the Ly $\alpha$ lines. They considered 18 QSO spectra and 42 metal line
systems and Ly $\alpha$ lines with rest equivalent width $ > \;
0.36\AA$ for their analysis and did not find any signal beyond 300 km
s$^{-1}$.

Since number of QSOs considered here is roughly 4 time more than that
used by Barcons and Webb (1990), and have uniform  signal to noise
ratio and resolution, it will be a worthwhile pursuit to reexamine the
cross-correlation properties.  The metal line systems used for  the
analysis are  given in table 1. Though the Ly $\alpha$ line list is
fairly homogeneous the metal line systems are inhomogeneous as they
were identified by various groups with spectra of varied resolution and
line selection criteria. In order to have a uniformity a grade is
assigned to each system using the criteria prescribed by York et al.
(1991).  If C IV or Mg II doublet was observed with the correct doublet
ratio, the system was assigned a grade 'B'. If in addition, plausible
lines of another species (besides Ly $\alpha$) were identified and
supported this doublet, the grade is assigned as 'A'. If doublet ratio
was inverted, or only one member of a doublet was identified, or, if a
spectrum was pictured in the reference and the lines did not appear at
all convincing, or the data was especially of poor resolution or of
poor signal-to-noise, then the grade assigned was 'C'.  Some of the
systems show broad and diffuse absorption lines of highly ionized
elements without strong Ly $\alpha$ absorption lines.  It is possible
that these lines are not  produced by the intervening clouds rather
they are due to ejected materials from the QSOs. These systems  are
rated as 'C' as the redshift of these systems need not reflect the
actual distances.  There are Lyman Limit systems in some of the spectra
used here which do not show any metal lines. Since it is not clear
whether they are very low metallicity metal line systems or  high
column density Ly $\alpha$ clouds these systems are not  considered for
the analysis. Thus in a way the possible uncertainties that may be
introduced in the cross-correlation  because of metal line system
selection criteria are avoided.

Only Ly $\alpha$ lines and the metal line systems which are 8 Mpc away
from the QSOs are considered to avoid any uncertainty due to proximity
effect.  Following Barcons and Webb (1990) in the case of multiple
system objects, the spectra  is splitted at the mid-point (in Ly
$\alpha$ redshift) between heavy element systems, and each spectral
segment is treated as the independent region of space. The
cross-correlation function ${\rm \xi_{ML}(v)}$ at different velocity
separations are calculated. The expected number in each bin is
calculated as described for pair-velocity correlation. Since the Ly
$\alpha$ line associated with metal line system obscures a significant
region of the adjacent Ly $\alpha$ forest, the interval ${\rm
\lambda_{obs} -W_{obs}/2,\; \rm \lambda_{obs}+W_{obs}/2}$ centered on
this observed Ly $\alpha$ line is excluded from the analysis.

The observed and expected number of Ly $\alpha$ lines in different
velocity bins are presented in figures 5a 5b and 5c, for ${\rm w_{min}
= 0.3\AA}$ for differently graded metal line systems. A glance at the
figure reveals that there is no excess in the lower velocity scales,
rather there seems to be a deficit of lines. This deficit seems to be
more that 1$\sigma$ level in the velocity interval 200-400 km s$^{-1}$
and consistent with the expected number within 1$\sigma$ for other
velocity bins. The deficit seems to be roughly 2$\sigma$ when only
systems graded as A are considered. As per our expectations the
difference between observed and expected number decreases as the
doubtful systems are also included in the analysis. Consistent with the
results of Barcons and Webb (1990), we are not finding strong
clustering signal in the lower velocity scales.  Thus it seems that the
Ly $\alpha$ clouds distribution, with equivalent width ${\rm >
0.3\AA}$, are fairly insensitive to the presence of metal line systems
beyond 200 km s$^{-1}$. Such a thing is expected in a biased CDM model
of structure formation, where the Ly $\alpha$ cloud distribution do not
follow the mass distribution.

\subsection{ Dependence of cross-correlation on the rest equivalent
width}

In the biased gravitationaly induced structure formation models, or if
the small fraction of Ly $\alpha$ clouds associate with metal line
systems one would expect to see the cross-correlation function to
depend on rest equivalent widths. Here we examine the possible
dependence of cross-correlation on the rest equivalent width of the Ly
$\alpha$ lines. We performed the analysis for ${\rm W_{min}}$ = 0.15,
0.30, 0.60 $\AA$ for differently graded metal line systems and the
obtained values of ${\rm \xi_{ML}(v)}$ are given in table 3. A glance
at the table 3 reveals no tendency of ${\xi_{\rm ML}}$ to depend on the
rest equivalent width of the Ly $\alpha$ clouds. Thus our results
indicate the distribution of strong as well as weak lines are fairly
uniform and not affected by the presence of metal line absorbers beyond
200 km s$^{-1}$ at higher redshifts.

\subsection { Dependence of cross correlation on redshift}

Recent HST observations presented by Bahcall et al. (1995) show the
clumps of Ly $\alpha$ lines  cluster around the metal line systems.
Lanzetta et al (1995) show at least 32$\%$ of the Ly $\alpha$ clouds
are associated with luminous galaxies at the very low redshifts. simple
extrapolation of these results demand the cross-correlation to increase
with time. If the Ly $\alpha$ clouds are formed via gravitational
instability in a biased scenario, the cross-correlation function is
expected to decrease with decreasing redshift due to the  environmental
dependent cloud evolution.  Here we investigate the possible dependence
of the cross-correlation on redshift .  The metal line systems with
grade A and B alone are considered for this analysis. The sample is
divided  into two at redshift z = 2.8, the expected and observed number
of pairs in various velocity bins are calculated for different values
of ${\rm W_{min}}$. The results are shown in figures 6a, 6b, 6c and
6d.

The observed correlation in the low velocity intervals are given in
table 4. In almost all cases we see a slight increase  in $\xi_{\rm
ML}$ for high z subsample. However the values are with in 1$\sigma$
errors and fairly consistent with no correlation. Note Barcons and
Webb(1990) also got similar trent( $\xi_{\rm ML} = -0.01\pm0.25$ for
low and 0.57$\pm$0.42 for high redshift sub-samples). However as noted
by them one requires more data to make any firmer statement on the
evolution of $\xi_{\rm ML}$.  If this trend is confirmed it will
provide a good means to probe the dynamical history of the Ly $\alpha$
clouds.

\section {DISCUSSION}

Our analysis confirms the existence of clustering in the low velocity
scales in the case of Ly $\alpha$ clouds. The redshift dependence of
the correlation function can discriminate between various models of Ly
$\alpha$ clouds. Though tentative results obtained in our analysis
indicate the increase of clustering with time, the data is noisy to
make any significant statement.

Sample used in this analysis is obtained with intermediate resolution
spectroscopic observations, and thus the lines observed are actually
blends of few lines. Barcons and Webb (1991) showed that blending of
randomly distributed lines alone cannot produce the observed equivalent
width distribution of the low resolution samples from the column
density distribution obtained using high resolution samples. They need
an enhanced blending due to clustering among Ly $\alpha$ clouds to
explain the observed results consistently. Thus our correlation
analysis will under predict the correlation amplitude in the smaller
scales ${\rm \le 200\; km\; s ^{-1}}$. The velocity interval considered
in this analysis, 200 - 600 km s$^{-1}$, will not be affected by
blending.  Travese, Giallongo {\&} Camurani(1992) showed that, using
the simulated spectra, the blending of lines can artificially increase
the number of high equivalent width lines more than the number of low
equivalent width lines and  produce an apparent differential evolution
in number density.  If the blending is severe in the sample used here
then the enhanced evolution in two point correlation shown by the high
redshift strong lines may be just an artifact of
blending and the real redshift evolution need not be the function of
equivalent width.  Only very large sample compiled with high resolution
spectra can give a better answer.

One can ascribe the lack of cross-correlation of Ly $\alpha$ lines with
metal line systems, as seen in our analysis, as the effect of
photoionization due to radiation from the metal line systems. Which
implies very high UV flux and hence very high star formation rate (SFR)
in the metal line absorbers. However the necessary SFR to shield the
cross-correlation excess can be ruled out based on the blank sky Ly
$\alpha$ emission searches (Srianand $\&$ Khare 1995a). Another
possibility may be the merger of Ly $\alpha$ lines into the metal line
systems. In this case one would like to see some time dependent
evolution of cross-correlation function.  Thus getting a statistically
significant estimate of redshift dependence of cross-correlation
function will enable us to understand the evolutionary scenarios.

If results obtained here are true then based on simple extrapolation
one should not find  the clustering of Ly $\alpha$ lines with luminous
galaxies at very low redshifts. However Lanzetta et al(1994) showed at
least 30 ${\%}$ of the Ly $\alpha$ clouds seem to associate with
luminous galaxies.  Thus these systems do not represent the population
of Ly $\alpha$ lines seen at higher redshifts if the metal line systems
seen at earlier epoch are progenitors of the present day galaxies.  Mo
and Morris (1994) showed that the very low redshift Ly $\alpha$
population can be a combination of mini halo population and the
absorbers associated with galaxies themselves.  Lanzetta et al(1994)
also showed the distribution of relative velocity between the Ly
$\alpha$ systems and the luminous galaxies are consistent with the
coincidence galaxies being  responsible for the absorption systems.
Thus these systems may be due to clouds in the extended halos around
the galaxies or due to extended disks (Maloney 1992) or tidel tails
(Morris and Van den Bergh; (1994)) produced due to interactions among
the galaxies. In such a scenario the observations with very high S/N
should show some associated metal lines in these systems

\end{document}